\documentclass[%
 reprint,
 amsmath,amssymb,
 aps,
]{revtex4-1}

\usepackage{graphicx}% Include figure files
\usepackage{dcolumn}% Align table columns on decimal point
\usepackage{bm}% bold math
\begin{document}

%\title{Atomic and electronic structure of A$_{x}$phenanthrene (A = Ca, Sr, Ba; x=1, 1.5, 2) and B$_{2}$phenanthrene (B = K, La)from a first principle study}
%\title{Atomic structure of Barium-doped phenanthrene and its electronic properties}
%\title{Crystal structure and electronic properties of Barium-doped phenanthrene}
\title{Ba$_2$phenanthrene is the main component in the Ba-doped phenanthrene sample}

\author{Xun-Wang Yan$^{1,2,3,4}$}
\email{yanxunwang@163.com}
\author{Zhongbing Huang$^{1,2}$}
\email{huangzb@hubu.edu.cn}
\author{Hai-Qing Lin$^1$}
%\author{Shaojing Qin$^3$}
\affiliation{$^1$Beijing Computational Science Research Center, Beijing 100084, China}
\affiliation{$^2$Faculty of Physics and Electronic Technology, Hubei University, Wuhan 430062, China}
\affiliation{$^3$State Key Laboratory of Theoretical Physics, Institute of Theoretical Physics,
Chinese Academy of Science, Beijing  100190, China }
\affiliation{$^4$School of physics and electrical engineering, Anyang Normal University, Henan 455000, China}
\date{\today}

\begin{abstract}
%For the newly discovered aromatic superconductor, there is a great challenge to determine the positions of metal atoms doped into molecular crystal whether in experiment or in theoretical simulation. In this paper,
We systematically investigate the crystal structure of Ba-doped phenanthrene with various Ba doping levels by the first principle calculation method combining with the X-ray diffraction (XRD) spectra simulation. The optimized lattice parameters and the simulated XRD spectra of Ba$_2$phenanthrene are in good agreement with the experiment ones, and the strength difference of a few of XRD peaks can be explained by the existence of undoped phenanthrene in the experimental sample. Although the stoichiometry ratio of Ba atom and phenanthrene molecule is 1.5:1, the simulated XRD spectra, space group symmetry and the optimized lattice parameters of Ba$_{1.5}$phenanthrene are not consistent to the measured values. The sample of Ba-doped phenanthrene is deduced to be the mixture of Ba$_2$phenanthrene and some amount of undoped phenanthrene, instead of uniform Ba$1.5$phenanthrene.
Our calculations indicate that Ba$_2$phenanthrene is a semiconductor with a small energy gap less than 0.05 eV. Our findings provide the fundamental information of crystal structure and electronic properties of Ba-doped phenanthrene superconductor.
%that the theoretical lattice parameters and inner atomic positions of Ba$_2$phenanthrene are well consistent to the experimental ones.
%The optimized lattice parameters of Ba$_2$phenanthrene are in perfect agreement with the experiment ones with the errors less than 1$\%$, and the simulated X-ray diffraction spectra indicate the doped Ba atoms position is consistent to experiment. Inner atomic positions, especially for Ba atoms, are also consistent to experiment by comparison of X-ray Diffraction peaks related to Ba atoms can be reproduced Mo

%Our calculations suggest that A$_2$phenanthrene (A = K, Ca, Sr, Ba or La) has a special dopant level and is the most probable parent compound for metal atom doped phenanthrene superconductor.
%The crystal structure of Ba-doped phenanthrene including the lattice parameters and inner atomic position is well reproduced by our calculation
%our results provide important crystal and electronic information for understanding of superconductivity in aromatic supercondutor.
\end{abstract}

\pacs{74.70.Kn, 74.20.Pq, 61.66.Hq, 61.48.-c}

\maketitle

%\section{Introduction}
%1 describe conducting molecular solid)
%The superconductivity in organic materials are among the most fascinating phenomena and have been a subject of considerable interest in condensed matter physics. Since the first organic superconductor (TMTSF)$_2$PF$_6$ was synthesized in 1979 \cite{Bechgaard19801119}, several classes of organic superconductors have been discovered, such as the quasi-one-dimensional (TMTSF)$_2$X \cite{IshiguroT}, two-dimensional (BEDT-TTF)$_2$X \cite{Urayama1988}, K-intercalated C$_{60}$\cite{Hebard1991}, Ca-doped graphite\cite{Emery2005}, in which TMTSF is tetramethyltetraselenafulvalene (C$_{10}$H$_{12}$Se$_4$), BEDT-TTF is bis(ethylenedithio)tetrathiafulvalene (C$_{10}$H$_8$S$_8$). The notable feature of organic superconductors is that there exits charge transfer between different two components in crystal and the carrier is delocalized over the the whole organic crystal by the $\pi$ molecular orbits, which is closely related to their superconductivity.

Recently, the discovery of superconductivity in potassium intercalated picene (C$_{22}$H$_{11}$) \cite{Mitsuhashi2010} by R. Mitsuhashi {\it et al.} in 2010 provides a new platform to explore the relationship of crystal structure, electronic property, magnetism and superconductivity.
%, because it is a new family of organic superconductor based on the aromatic hydrocarbons.
Subsequently, phenantherene (C$_{14}$H$_{10}$) \cite{Wang2011} and dibenzopentacene (C$_{30}$H$_{18}$)\cite{Xue2012} were intercalated by potassium to synthesize new superconductors, %which were composed of fused benzene rings and
which had similar molecular structure to picene just with the different number of benzene rings.
%Another approach to synthesize new aromatic superconductor is that
On the other hand, alkaline earth and rare earth metal in place of alkaline metal were adopted to intercalated into phenanthrene crystal to explore their superconductivity. In experiment, the high quality Ba$_{1.5}$phenanthrene\cite{Wang2011a} and La$_1$phenanthrene \cite{Wang2012} superconductors are obtained with superconductive shielding fractions $40\%$ and $46\%$.

Although some new progresses on the aromatic superconductors have been made in experiment, such as new synthesis method for K$_3$picene \cite{Kambe2012}, confirmation of superconductivity from resistivity measurement \cite{Teranishi2013} and pressure effect on superconductivity\cite{Chen2013,Kambe2012}, there still remains lots of uncertainty in this new class of organic superconductor.
Firstly, the detailed crystal structures of doped aromatic molecular solids have not yet been reported in experiment, especially the metal atom position, due to the sample degradation in air and the limit of measurement techniques \cite{Wang2012}.
And for the previous theoretical simulation on the crystal structure, the optimized lattice parameters for metal doped picene and phenantherene have large discrepancy to the experimental ones \cite{Kosugi2011,DeAndres2011a,DeAndres2011,PhysRevB.88.115106,Yan2013}.
Secondly, the importance of electronic correlation and magnetism in metal doped aromatic solid were emphasized by several research groups \cite{Kim2011,Giovannetti2011,Kim2013a,Huang2012,Verges2012} and they thought that their superconductivity is unconventional, but other groups thought the mechanism of superconductivity in aromatic superconductors could be explained in the framework of Bardeen-Cooper-Schrieffer (BCS) theory \cite{Bardeen1957} \cite{Kato2011}.
Thirdly, whether metal or insulator is the parent compounds of aromatic superconductor is in debat.
Many experimental and theoretical researches think charge transferring from metal to molecular unoccupied orbits lead to metallicity and further to superconductivity in the class of compounds. However, Andreas Ruff {\it et al.} investigated the absence of metallicity in K-doped picene film on the Si substrate by photoelectron spectroscopy and {\it ab~initio} density functional theory combined with dynamical mean-field theory (DFT + DMFT) \cite{Ruff2013}. The insulating K-doped picene on the Au substrate \cite{Caputo2012}and insulating La-doped phenanthrene \cite{PhysRevB.88.115106,Yan2013} were also reported in experiment and theoretical simulation.

%on the most details related to the precise chemical composition, crystal structure, the role of electronic correlation to superconductivity, or even whether metal or insulator for the parent compound.
But anyway, it is noticed that the determination of the atomic structure for metal doped aromatic crystal is the first step to explore the electronic properties and the superconducting mechanism. Unfortunately, the simulated lattice parameters in the recent theoretical researches on the K- or La-intercalated phenanthrene or picene do not reproduce the experimental lattice.
In this paper, we focus on alkaline earth metal barium doped phenanthrene, for which there is no theoretical exploration on the crystal structure and its electronic structure remains unknown.
%And, we found out the crystal structure
%of Ba$_2$phenanthrene which lattice parameters are in perfect agreement with the experiment ones.
We perform the optimization of the crystal structure of Ba-doped phenanthrene under different doped concentration, and find that the lattice parameters of Ba$_2$phenanthrene are in perfect agreement with the measurements. The simulated XRD spectra of Ba$_2$phenanthrene indicate the doped Ba atom positions are also consistent to the experiment. But the case of Ba$_{1.5}$phenanthrene is not good as Ba$_2$phenanthrene. We can conclude that the experimental sample is mainly composed of Ba$_2$phenanthrene mixed with some pristine phenanthrene, instead of unifrom Ba$_{1.5}$phenanthrene.

%The paper is organized as the following order. First, the optimized structures for Ba$_1$phenanthrene, Ba$_{1.5}$phenanthrene, Ba$_2$phenanthrene are presented; Then the electronic structure of Ba$_{1.5}$phenanthrene, Ba$_2$phenanthrene are investigated; Last we discuss the components of the experimental sample with the simulated XRD spectra.

%\section{Method and details}
In our calculations the generalized gradient approximation
(GGA) with Perdew-Burke-Ernzerhof (PBE) formula~\cite{PhysRevLett.77.3865} was adopted for the exchange-correlation potentials. The projector augmented-wave method (PAW)
pseudopotentials were used to model the electron-ion interactions \cite{PhysRevB.50.17953}.
The C, H, Ba psudopotentials are from the subfolder C$\_s$, H$\_s$, Ba$\_{sv}$, K$\_{sv}$, La in the pseudopotential package potpaw$\_$PBE.52 supplied by Vienna Ab initio simulation package (VASP) website \cite{PhysRevB.47.558, PhysRevB.54.11169}. The plane wave basis cutoff is set to 300 eV.
%and it is sufficient since the default cutoff energy maximum in the pseudopotential files used in our calculation is less than 260 eV.
The Gaussian broadening technique was
used and a mesh of $4\times 4\times 4$ k-points were sampled for the Brillouin-zone integration.
%In the calculations, the lattice parameters with the internal atomic coordinates were optimized by the energy minimization.
The convergence thresholds of the total energy, force on atom and pressure on cell are 10$^{-4}$ eV, 0.01 eV/\AA ~and 0.1 KBar respectively, which criteria are all satisfied in calculation.
%In order to check our results, van der Waals density functional calculations have also been performed. The nonlocal dispersion interaction was considered in the scheme developed by Dion, Thonhauser {\it et. al } \cite{PhysRevLett.92.246401,PhysRevB.83.195131} enclosed in first principle code VASP.

% emphasize the Ba-doped alkali earth metal doped is not report and the electronic structure is not clear.
%, emphasize the challenge for experiment and simulation, there are a great difficulty, for example , there are not found out the correct structure.
% emphasize the importance of correct crystal structure. is a basis for all understanding.
% we found the crystal structure,
%van der waals interaction, should be included in the study.
% emphasize K2 K3 experiment, picene K2 K3 is the compound.
% refer to the K1 K2 K3 is the insulating phases.
%1 describe three experiments + chenxianghui Ba La experiment)
%2 describe the electronic structure papers  , theory)
%(3 point out var de waals interaction importance)
%4 refer to this paper and this paper goal)
\begin{figure}.
\includegraphics[width=8.0cm]{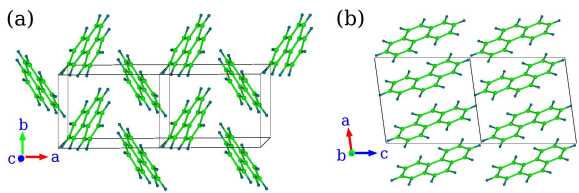}
\caption{(Color online)
The intralayer arrangement of molecules is shown in a 2 $\times$ 2 $\times$ 1 supercell (a) and the interlayer arrangement of molecules is shown in a 1 $\times$ 1 $\times$ 2 supercell (b) for pristine phenanthrene. There are two phenanthrene molecules in a unit cell.} \label{pristine-struct}
\end{figure}

\begin{figure}.
\includegraphics[width=8.0cm]{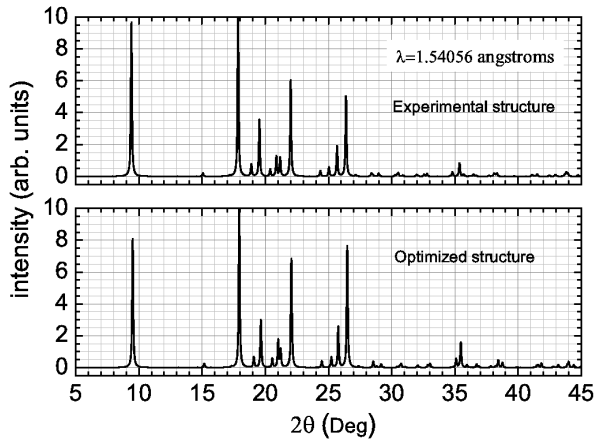}
\caption{(Color online)
The simulated XRD spectra of pristine phenanthrene based on the experimental crystal structure (a) and the optimized structure (b). The experimental crystal structure is from Ref.\cite{Kay1971}, and the used X-ray wave length is 1.54056 \AA.} \label{pristine-XRD}
\end{figure}

%(7)(display the main result)
%describe the crystal structure.
Pristine phenanthrene is a typical kind of molecular solid and crystallizes in the space group $P_{21}$. Each unit cell contains two molecules which are arranged with a herringbone pattern, and phenanthrene molecules form a layer parallel to $ab$ plane and then molecular layers are stacked along the $c$ axis. The crystal geometry is shown in Fig. \ref{pristine-struct}.
We first simulate the crystal structure of pristine phenanthrene, which can examine the applicability of C and H pseudopotentials in our calculation and can provide a reference for exploring the structure of the Ba-doped phenanthrene.
After cell parameters and inner atom positions relaxation, we obtain the lattice parameters of pristine phenanthrene $a = 8.445 $ \AA, $b = 6.118$ \AA, $c = 9.372 $ \AA ~and $\beta = 97.81 ^{\circ}$, in a perfect agreement with the experimental parameters $a = 8.46 $ \AA, $b = 6.16$ \AA, $c = 9.47 $ \AA~ and $\beta = 97.7 ^{\circ}$ \cite{Kay1971}. For the experimental structure and our optimized structure of phenanthrene, we have calculated the XRD sepectra with Mercury program \cite{mercury}, shown in Fig. \ref{pristine-XRD}. The consistency between two spectra demonstrate that the molecular positions in the cell in our calculation are in accord with the experiment measurement.

%\subsection{Ba-doped phenanthrene }
%%transition sentence
Although phenanthrene and Ba power are mixed with a stoichiometric ratio of $1:1.5$ in experiment, for the synthesized Ba$_x$phenanthrene there are several possible values of $x$, such as $1$, $1.5$, $2$, due to structural phase separation or non-uniform mixing in practice. So we explore the crystal structure of Ba$_x$phenanthrene with $x = 1, 1.5, 2$.

\begin{table*}
\caption{\label{Ba-table} The optimized lattice parameters $a, b , c, \alpha, \beta, \gamma$ and the space group of unit cell for Ba-doped phenanthrene with various Ba concentration. The difference of parameters from the experimental values are listed in brackets. The length unit is \AA~ and angle unit is degree.}

\begin{tabular}{l c c c c c c c}
\hline
\hline
         &a &b &c & $\alpha$ & $\beta$ & $\gamma$ & space group \\
\hline
Ba$_1$phenanthrene(A)     &8.81(+0.32) &6.19(+0.01) &9.50(+0.0)  &89.48 &99.77  & 87.61& P1\\
Ba$_1$phenanthrene(B)     &8.45(-0.03) &6.54(+0.36) &9.41(-0.09) &90.0  &102.45 &90.0 & P2$_1$\\
%Ba$_{1.5}$phenanthrene(A) &9.39(+0.91) &6.56(+0.38) &9.52(+0.02) &90.04 &101.24 & 94.08& P1 \\
Ba$_{1.5}$phenanthrene &8.43(-0.05) &6.68(+0.50) &9.40(-0.10) &88.97 &104.26 & 88.42& P1\\
%Ba$_{2.0}$phenanthrene(A) &9.07(+0.59) &6.45(+0.27) &9.50(+0.0)  &88.58 &99.23  &92.79 & P1\\
%Ba$_{2.0}$phenanthrene(B) &8.89(+0.41) &6.31(+0.13) &9.76(+0.26) &84.82 &100.31 &90.94 & P1\\
Ba$_{2.0}$phenanthrene &8.64(+0.16) &6.40(+0.22) &9.76(+0.26) &90.00 &102.55 &90.00 & P2$_1$\\
Experiment\cite{Wang2011a}                      & 8.48       &6.18        &9.50        &90.0  & 97.49 &90.0& P2$_1$ \\
\hline
\hline
\end{tabular}
\end{table*}

\begin{figure}.
\includegraphics[width=8.0cm]{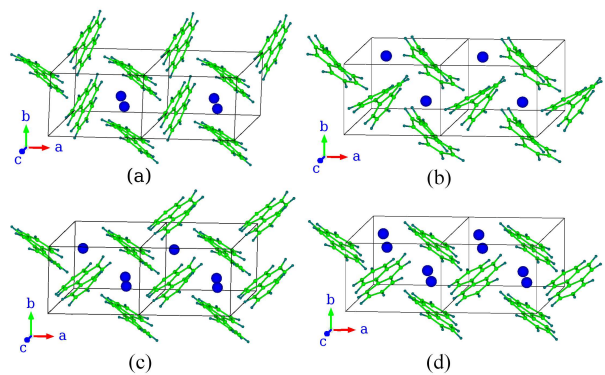}
\caption{(Color online)
Schematic structure of the 2 $\times$ 1 $\times$ 1 supercell for Ba$_1$phenanthrene(A), Ba$_1$phenanthrene(B), Ba$_{1.5}$phenanthrene and Ba$_2$phenanthrene, corresponding to (a), (b), (c) and (d), respectively. The unit cell shown in (a), (b), (c) or (d) is composed of 2, 2, 3 or 4 Ba atoms and two phenanthrene molecules. Large blue balls represent Ba atoms.} \label{2Ba-struct}
\end{figure}

%\begin{figure}.
%\includegraphics[width=8.5cm]{3Ba-sit-twosides.eps}
%\caption{(Color online)
%The intralayer arrangement of molecules is shown in a 2 $\times$ 2 $\times$ 1 supercell (a) and the interlayer arrangement of molecules is shown in a 1 $\times$ 1 $\times$ 2 supercell (b) for pristine phenanthrene. There are two phenanthrene molecules in a unit cell.} \label{3Ba-struct}
%\end{figure}
%\begin{figure}.
%\includegraphics[width=14cm]{4Ba-sit-twosides.eps}
%\caption{(Color online)
%The intralayer arrangement of molecules is shown in a 2 $\times$ 2 $\times$ 1 supercell (a) and the interlayer arrangement of molecules is shown in a 1 $\times$ 1 $\times$ 2 supercell (b) for pristine phenanthrene. There are two phenanthrene molecules in a unit cell.} \label{4Ba-struct}
%\end{figure}

In a molecular layer of phenanthrene crystal, phenanthrene molecules are arranged in a herringbone manner and the regular interstitial spaces are formed. We call the interstitial space as hole and can find that each unit cell have two holes averagely.
According to the previous studies \cite{Kosugi2009,Verges2012}, the dopants are accommodated in the intralayer region of the picene and phenanthrene solids, instead of the interlayer region.
A Ba$_1$phenanthrene unit cell consists of two Ba atoms and two molecules.
There are two possible geometry configurations, one is that two Ba atoms sit in a hole with the neighbor hole unoccupied, another is that each hole has a Ba atom, called them as Ba$_1$phenanthrene(A) and Ba$_1$phenanthrene(B) respectively, shown in the Fig.\ref{2Ba-struct}(a) and (b).
For Ba$_{1.5}$phenanthrene or Ba$_2$phenanthrene, the unit cell is composed of three or four Ba atoms, which distribution in two holes is 1:2 or 2:2, see Fig.\ref{2Ba-struct} (c) and (d). Other some arrangements of Ba atoms with more than two atoms in a hole are also considered (not shown). For the four structural phases, the optimized lattice parameters and space group symmetry are listed in the Table.\ref{Ba-table}.

Firstly, we exclude the Ba$_1$phenanthrene(A) and Ba$_1$phenanthrene(B) and think them to be non-real structure.
For Ba$_1$phenanthrene(A), its energy per unit cell decrease 0.18 eV relative to Ba$_1$phenanthrene(B), which means Ba$_1$phenanthrene(A) is more stable structure when the ratio of phenanthrene molecule and Ba atom is $1:1$. The symmetry of Ba$_1$phenanthrene(A) is P1 space group, instead of the measured symmetry $P_{21}$, and the parameter $\gamma$ = 87.61 \AA~ leads to the cell shape deviating from the monoclinic cell. So, we exclude the Ba$_1$phenanthrene(B) in terms of energy and exclude Ba$_1$phenanthrene(A) in terms of symmetry and lattice parameters.

Secondly, we check the case of Ba$_{1.5}$phenanthrene. Its unit cell is composed of three Ba atoms and two molecules. Because the number of Ba atoms is odd, there is no equivalent atom for the third Ba atom in terms of the symmetry operation of P2$_1$ space group ($x,y,z \rightarrow -x,y+0.5,-z$). Therefore, Ba$_{1.5}$phenanthrene can not crystalize in space group P2$_1$. As can be seen from Table.\ref{Ba-table}, $b$, $\alpha$ and $\gamma$ have a large discrepancy to the experimental values.
We also examine the energy change in the process that one metal atom is added into Ba$_{1.5}$phenanthrene to form Ba$_{2}$phenanthrene. The formation energy is defined as $E_{formation(1.5\rightarrow2)} = E_{2}-E_{1.5}-E_{metal}/n$. $E_{2}$ is the energy of unit cell for Ba$_{2}$phenanthrene, $E_{1.5}$ is the energy of unit cell for Ba$_{1.5}$phenanthrene and $E_{metal}/n$ is the energy of one metal atom in elemental metal. The $E_{formation(1.5\rightarrow2)}$ is -0.21 eV. The negative values tell us that the process is easy to occur. So, we conclude that Ba$_{1.5}$phenanthrene is also not the real structure in experiment.

\begin{figure}.
\includegraphics[width=8.0cm]{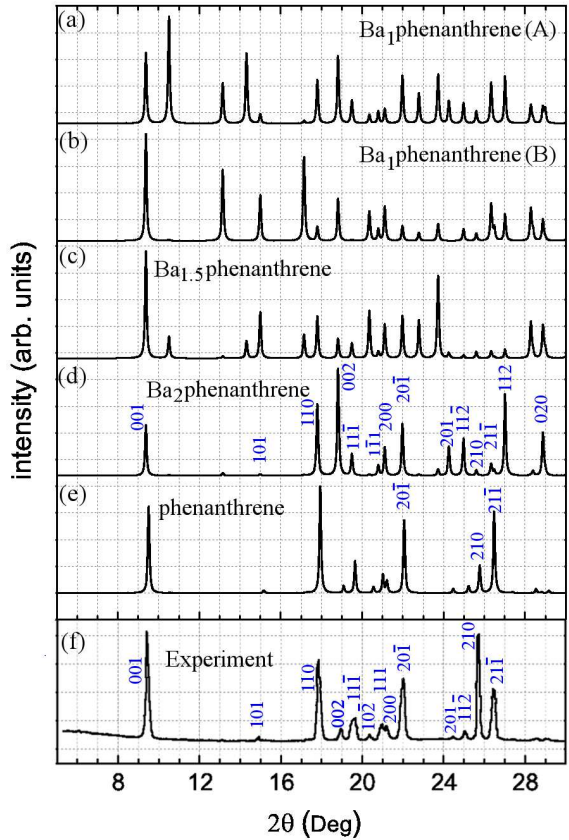}
\caption{(Color online)
The simulated XRD spectra for Ba$_1$phenanthrene(A), Ba$_1$phenanthrene(B), Ba$_{1.5}$phenanthrene, Ba$_{2}$phenanthrene and pristine phenanthrene are presented in (a), (b), (c), (d), and (e). The experimental XRD spectra is shown in (f) and the data is from Ref.\cite{Wang2011a}. The peaks are indexed with the blue numbers. The used X-ray wave length is 1.54056 \AA.} \label{XRD-all}
\end{figure}

Lastly, we demonstrate that the arrangement of molecules and Ba atoms in Ba$_2$phenanthrene is the most reasonable structure.
Comparing to the measured values, the calculated parameter $a$, $b$ and $c$ for Ba$_2$phenanthrene increase 0.16 \AA, 0.22 \AA and 0.26 \AA. It is noticed that the discrepancies for three parameters are uniform, while only one of three parameters has a sharp increase relative to the experiment for Ba$_1$phenanthrene(A), Ba$_1$phenanthrene(B) and Ba$_{1.5}$phenanthrene, which can be seen in Table.\ref{Ba-table}.
In Ba$_{2}$phenanthrene, phenanthrene molecules have less distortion than those in Ba$_1$phenanthrene(B), though both Ba$_{2}$phenanthrene and Ba$_1$phenanthrene(B) hold the P2$_1$ space group symmetry in accord with the measurement. As mentioned above, the energy of Ba$_1$phenanthrene(A) is lower than Ba$_1$phenanthrene(B) and the formation energy $E_{formation(1.5\rightarrow2)}$ is negative. These facts indicate definitely that the configuration of one hole accommodating two Ba atoms is the stable structure.
see Fig. \ref{XRD-all}.

In order to further confirm our judgement on the Ba atom arrangement in Ba-doped phenanthrene,
we investigate the XRD spectra of Ba-doped phenanthrene.
The peaks in XRD spectra correspond to the different crystal planes, which can directly reflect atomic positions in crystal.
We fix the lattice parameters at the experimental values and optimize atomic positions in the unit cell for the above four structure configurations.
The simulated XRD spectra are presented in Fig. \ref{XRD-all}.
At first, we compare the simulated XRD spectra of Ba$_1$phenanthrene(A), Ba$_1$phenanthrene(B) and Ba$_{1.5}$phenanthrene
with the measured XRD spectra of Ba-doped phenanthrene, shown in Fig. \ref{XRD-all} (a), (b), (c) and (f).
The profiles of three XRD spectra have a great discrepancy to the measured ones. Specifically, there exist a few of strong peaks in the degree range 10$^{\circ}$ - 17 $^{\circ}$ and 22$^{\circ}$ - 25$^{\circ}$ in the three simulated XRD spectra, but no the corresponding peaks in the measured spectra. And the strength of several main peaks, for example (100), (20$\bar{1}$) and (210), are obviously less the experimental peaks. These extra peaks and differences of peak strength indicate that the crystal structures for Ba$_1$phenanthrene(A), Ba$_1$phenanthrene(B) and Ba$_{1.5}$phenanthrene do not match the real structure.

\begin{figure}.
\includegraphics[width=8.5cm]{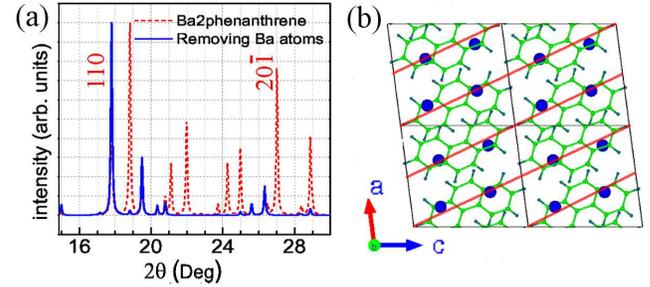}
\caption{(Color online)
(a), the simulated XRD spectra for Ba$_2$phenanthrene (blue line) and Ba$_2$phenanthrene but removing Ba atoms (red dash line);
(b), the schematic view of $2 \times 1 \times 2$ Ba$_2$phenanthrene supercell along $b$ axis, the red parallel lines represent the (20$\bar{1}$) crystal planes.} \label{XRD-deletBa}
\end{figure}

Then we focus on the XRD spectra of Ba$_2$phenanthrene, which is shown in Fig. \ref{XRD-all} (d) and the peak indexes are marked with the blue numbers. Unlike the above three structure configurations, its spectra is similar to the experimental XRD one. The peak positions in the degree range 5$^{\circ}$ - 27$^{\circ}$ are exactly consistent to the ones in the spectra in Fig.\ref{XRD-all} (f). Except for (002) peak, the strength of those peaks less than 24$^\circ$ including the little (101), (11$\bar{1}$), (1$\bar{1}$1) and (20$\bar{1}$) peaks, has a good agreement with the experimental spectra.
These peaks are in a tight correlation with Ba atoms positions. When we remove Ba atoms from the cell, the spectra is changed greatly, see Fig. \ref{XRD-deletBa} (a) (for direct view, the crystal planes corresponding to (20$\bar{1}$) peak is plotted in Fig. \ref{XRD-deletBa} (b)).
%Ba atoms positions are in a tight correlation with these peaks, which can be explained by Fig.\ref{XRD-deletBa}. Fig. \ref{XRD-deletBa} (a) shows that the spectra is changed greatly when Ba atoms are removed with the molecule positions unchanged and Fig. \ref{XRD-deletBa} (b) shows the crystal planes corresponding to (20$\bar{1}$) peak.
Generally speaking, the XRD spectra of Ba$_2$phenanthrene can reflect the main features of experimental XRD spectra.

Finally, how to explain the difference of (002), (210), (21$\bar{1}$) and other peaks in Fig. \ref{XRD-all} (d) and (f). Let us look back the experiment where the stoichiometry ratio of molecule and Ba atom is 1:1.5. According to our above analysis, Ba$_{2}$phenanthrene is the most possible structure phase with the ratio 1:2. Hence, there should exist residual undoped phenanthrene in the sample with about one forth of total phenanthrene. As displayed in Fig. \ref{XRD-all} (e), the XRD spectra of pristine phenanthrene resembles surprisingly the spectra in the Fig. \ref{XRD-all} (f). Therefore, the existence of quite a bit of phenanthrene can increase the strength of (210), (21$\bar{1}$) peak and decrease the strength of (002), (112) and (020) peak, make the XRD spectra of Ba$_2$phenanthrene more close to the measured XRD spectra.

%To explain this correlation, we remove Ba atoms from the unit cell and keep the molecule positions unchanged to calculate the XRD sepctra. After removing Ba atoms, the spectra is changed greatly and some main peaks disappeare, see Fig. \ref{XRD-deletBa} (a). We take (20$\bar{1}$) peak as an example to present its related crystal plane shown in Fig.\ref{XRD-deletBa} (b).

\begin{figure}
\includegraphics[width=8.0cm]{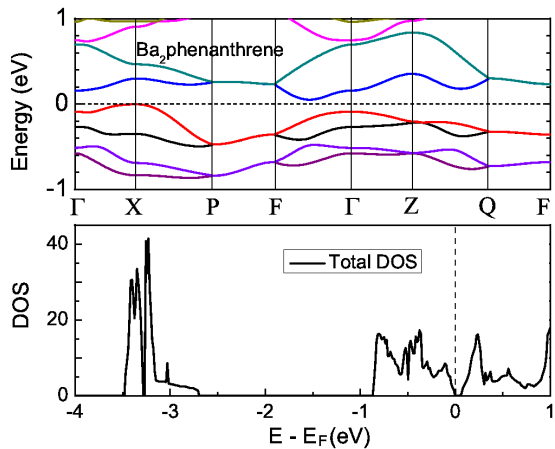}
\caption{(Color online)
Top panel: The band structure of Ba$_2$phenanthrene based on the optimized structure. The $k$ point path is the same to Fig.4 in Ref.\cite{Yan2013}. Bottom panel: The total density of state of Ba$_2$phenanthrene. Fermi level is set to zero.} \label{dos-band}
\end{figure}

%%%\subsection{The electronic structures of A$_{2}$phenanthrene (A = Ba, Sr, K, La)}
We further perform the first principle electronic structure calculation for Ba$_{2}$phenanthrene. The band structure and the density of state (DOS) based on the fully optimized structure are presented in Fig.\ref{dos-band}. Ba$_{2}$phenanthrene is a semiconductor with a very small energy gap less than 0.05 eV at the Fermi level. Two pairs of bands below Fermi energy result from two pairs of molecular orbitals - the lowest unoccupied molecular orbital (LUMO) and LUMO+1, related to two molecules in a unit cell. Four Ba atoms in a unit cell provide eight electrons to occupy the two pairs of energy bands. The bands related to the highest occupied molecular orbitals (HOMO) locate at -2.7 eV shown in the bottom panel of Fig.\ref{dos-band}, and the gap from -2.7 eV to -0.9 eV corresponds to the energy gap at Fermi level for pristine phenanthrene.
%On the whole, the bands around Fermi energy are flat and little dispersive, which lead to no overlap or very small overlap between two pairs of band, corresponding to the deep 'valley' in the DOS spectra.
%For A$_{2}$phenanthrene (A = Ba, Sr, K, La), the measured superconducting transition temperature T$_c$ are all about 5 K. The similarity of T$_c$ values maybe relate to the narrow energy gap and deep DOS 'valley' at Fermi level -- the common feature of electronic structure for these compounds.
As to the superconductivity in Ba-doped phenanthrene, Ba$_{2}$phenanthrene should be the structural phase of parent compound. When the concentration of Ba has few percents
deviation from 2 in Ba$_{2}$phenanthrene, similar to the parent compounds of superconductors LaCuO4, the carriers can appear to lead to the metallicity and superconductivity.

In conclusion, the first principles investigation for the crystal structure and electronic properties of Ba-doped phenanthrene with various Ba concentrations have been performed. By comparing the simulated lattice parameter, space group symmetry with the experimental ones, we find that the structural phase of Ba$_{2}$phenanthrene have a good agreement with the experimental structure. The simulated XRD spectra of Ba$_{2}$phenanthrene is also similar to the measured spectra, and the difference of a few of peaks can be explained by the existence of some amount of undoped phenanthrene in the sample. The negative formation energy prove that it is easy to form Ba$_{2}$phenanthrene. The electronic properties calculation indicate that Ba$_{2}$phenanthrene is a semiconductor with a small energy gap less than 0.05 eV. Although the stoichiometry ratio of Ba atom and molecule is 1.5:1, its lattice parameters, space symmetry, XRD spectra are not consistent with the experimental ones. Hence, we conclude that the real structural phase of Ba-doped phenanthrene in experiment is not uniform Ba$_{1.5}$phenanthrene, but Ba$_{1.5}$phenanthrene mixed with some amount of undoped phenanthrene.
To our knowledge, this is the first time that the simulated structure is in such a good agreement with the experiment and the first time that the positions for the dopant of metal atom are distinguished definitely for the metal doped aromatic superconductors.
Our findings clarify that Ba$_{2}$phenanthrene is the main component for Ba-doped phenanthrene and undoped phenanthrene also exist in the sample, which provide the important information of crystal structure and electronic properties of Ba-doped phenanthrene superconductor.

%\section{Acknowledgments}
Acknowledgments: We acknowledge Cai-Zhuang Wang for interesting suggestion and fruitful discussion, and thank Jun-Feng Gao and Xiaoli Wang for their help in XRD simulation. This work was supported
by MOST 2011CB922200, the Natural Science Foundation of China under Grants
Nos. 91221103, 11174072 and U1204108.
%, and was also partially supported by the program of the State Key Laboratory of Theoretical Physics (No.Y3KF271CJ1), Institute of Theoretical Physics, Chinese Academy of Sciences.

%\begin{references}
%\bibitem{Nature2010}
%Nature 464, 76-79 (4 March 2010)
%\bibitem{feng-naturem}
% Zhang Y, {\it et al.}, Nature Materials {\bf 10}, 273 (2011).
%\end{references}

\bibliographystyle{apsrev4-1}
%\bibliography{11,Bechgaard,Collection,achs_jpcbfk102_498,espresso,achs_jacsat130_3296,LaCuO4}
\bibliography{Collection20140305}

\end{document}